# Knowledge Integration and Diffusion:
# Measures and Mapping of Diversity and Coherence

Ismael Rafols *(i.rafols@ingenio.upv.es)*

*Ingenio (CSIC-UPV)*, Universitat Politècnica de València, València (Spain) &
SPRU - Science and Technology Policy Research University of Sussex, Brighton (England)

Chapter for:

Ying Ding, Ronald Rousseau, Wolfram Dietmar (Editors)
*Measuring scholarly Impact: Methods and Practice*

## Abstract

In this chapter, I present a framework based on the concepts of diversity and coherence for the analysis of knowledge integration and diffusion. Visualisations that help understand insights gained are also introduced. The key novelty offered by this framework compared to previous approaches is the inclusion of cognitive distance (or proximity) between the categories that characterise the body of knowledge under study. I briefly discuss the different methods to map the cognitive dimension.

**1. Introduction**

Most knowledge builds on previous knowledge –given the cumulative nature of science and technology. But the fact that knowledge mainly draws on previous knowledge also means that it does not build on "other" types of knowledge. This is what in an evolutionary understanding of science is called a cognitive trajectory -which is often associated with lock-in.[1] Under such conditions, the combination of different types of knowledge (perspectives, but also data, tools, etcetera) has long been seen as a way to leap out of stagnation and create new knowledge. This perspective has been emphasised in the case of research aiming to solve social and economic problems -- seen as requiring interdisciplinary efforts, both in terms of sources (i.e. requiring the integration of different types of knowledge) and it terms of impacts (i.e. diffusion over different areas of research and practice) (Lowe and Phillipson, 2006; Nightingale and Scott, 2007).

Changes in science in the last two decades have been characterised as a progressive blurring of the well defined categories of postwar science. Science has shifted towards a so-called Mode 2 of knowledge production that is presented as more interdisciplinary, more heterogeneous, closer to social actors and contexts, and more susceptible to social critique (Gibbons et al., 1994; Hessels and van Lente, 2008).

---

[1] It should be noted that the evolutionary view of science and technology is prevalent both among constructivist sociologists such as Bijker (Pinch and Bijker, 1984) and positivist economist such as Dosi (1982).

In Mode 2 research, knowledge integration and diffusion play a crucial role as the processes that bridge the gaps between disciplines, organisations, institutions and stakeholders. Building on Boschma's notion of multiple dimensions of proximity (Boschma, 2005), Frenken et al. (2010) proposed to:

> 'reformulate the concept of Mode 2 knowledge production analytically as a mode of distributed knowledge production, where we operationalize the notion of distribution in five proximity dimensions [i.e. cognitive, organisational, social, institutional, geographical] (...) Mode 1 stands for scientific knowledge production in which actors are distributed, yet proximate, while Mode 2 knowledge production stands for distributed knowledge production processes, in which actors are distant.'

While cognitive proximity is the primary dimension to analyse knowledge integration and diffusion in science, it is worth realising that other dimensions of proximity mediate the possibility of knowledge integration and diffusion.[2] These other dimensions are important to understand how changes in cognitive proximity happen. Policy and management instruments such as personnel recruitment, organisational reforms or incentives directly address these others dimensions (social, organisational or institutional) and it is through them that decision makers aim to influence the cognitive dimension. The Triple Helix framework, for example, investigates the institutional-cognitive-organisational relations (Etzkowitz and Leydesdorff, 2000). One can study 'translational research institutes', which increase geographical and organisational proximity between, for example a cell biologist and an oncologist, as efforts to favour integration and diffusion of knowledge between basic research and practice related to cancer (Molas-Gallart et al., 2013).

In this chapter, I review quantitative methods and some visualisation techniques developed in recent years in order to assess where, how and to which extent knowledge integration and diffusion took place regarding specific organisation, problem-solving efforts or technologies. While this chapter focuses on mapping of the cognitive dimension, I invite the reader to think, following Boschma and Frenken's proposal, that the understanding of the dynamics of science consists in being able to relate the different analytical dimensions.

## 2. Conceptual framework: knowledge integration and diffusion as shifts in cognitive diversity and coherence[3]

Let me start by defining knowledge integration as the process bringing into relation bodies of knowledge or research practice hitherto unrelated or distant. Similarly, I define knowledge diffusion as the movement or translation of a piece of knowledge to bodies of knowledge where it had not been used before. These 'specialised bodies' of knowledge can refer to perspectives, concepts, theories but also to tools, techniques or information and data sources (National Academies, 2004). For example, some of the key contributions of the very successful US National Center for Ecological Analysis and Synthesis (NCEAS) in UCSB were based precisely on cross-fertilisation of methods and data sources used in different fields within ecology (Hackett et al., 2008).

---

[2] The use of these five dimensions is an expedient simplification. One may easily conceive more dimensions within each of the dimensions listed making a more fine-grained description of cognitive or social dimensions, for example.
[3] This framework was first introduced in Rafols and Meyer (2010), then re-presented in more general form in Liu et al. (2012) and again in an empirical case in Rafols et al. (2012) with some substantial changes. Here I try to make a further generalisation of the concept of coherence in the hope of incremental improvements.

The difference between integration and diffusion is mainly one of perspective. For example, from the perspective of a Valencian laboratory working on breast cancer, RNA interference (RNAi) is *integrated* to their portfolio of methods for genetic manipulation, i.e. a piece of knowledge is integrated into the knowledge base of an organisation. However, from the perspective of an emergent technology such as RNAi, it is the RNAi technique which has *diffused* into a laboratory --a laboratory which is a point in a space that may be characterise by geography (València), discipline (oncology) or research problem (breast cancer). In this chapter, the emphasis is given to integration, but the frameworks proposed and many of the tools used can be used as well to analyse knowledge diffusion (Carley and Porter, 2012).

Both integration and diffusion are dynamic processes and, therefore, they should be analysed over time (Liu and Rousseau, 2010; Leydesdorff and Rafols, 2011a). It is, nevertheless also possible to make a static comparison of the degree of integration represented in different entities such as publications (Porter and Rafols, 2009), researchers (Porter et al., 2007) or university departments (Rafols, Leydesdorff et al., 2012).

The framework proposed here analyses separately the two key concepts necessary for the definition of knowledge integration. On the one hand, *diversity* describes the differences in the bodies of knowledge that are integrated, and on the other hand, *coherence* describes the intensities of the relations between these bodies of knowledge. Notice that the concept of diversity is interpreted in the same way in the case of integration and of diffusion. However, for coherence the interpretation differs for integration and diffusion. More coherence can be interpreted as an increase in integration (because knowledge has become more related). In the case of diffusion, more coherence does not mean necessarily more diffusion, but a specific type of diffusion: spread over topics in which these topics have become related.

Another way of studying knowledge integration (and interdisciplinarity) is to focus on the bridging role, or *intermediation* role of some specific scientific contributions, typically using notions from social network analysis such as betweenness centrality (Leydesdorff, 2007; Chen et al., 2009). In Rafols, Leydesdorff et al. (2012), we developed intermediation as a framework, complementary to diversity and coherence, which is useful to explore fine-grained, bottom-up perspectives of dynamics. However, for lack of space and expertise, I will leave intermediation outside of the scope of this chapter.

Given that integration can be analysed at different levels, let us first make a rather abstract description of diversity and coherence. We will consider the *system* or unit of analysis (e.g. university department), the *elements* (e.g. articles), the *categories* (e.g. Web of Science (WoS) categories) and the *relations* between categories (e.g. citations from one WoS category to another).

Diversity is a 'property of the apportioning of elements or options in any system' (Stirling, 1998, 2007, p. 709). For example, the disciplinary diversity of a university (system) can be proxied by the distribution of the articles (elements) published in WoS categories (categories) (as shown in Figure 1). Diversity can have three distinct attributes as illustrated in Figure 2:

- *variety*: number of categories into which the elements are apportioned (*N*).
- *balance*: evenness of the distribution of elements across categories.
- *disparity*: degree to which the categories of the elements are different.

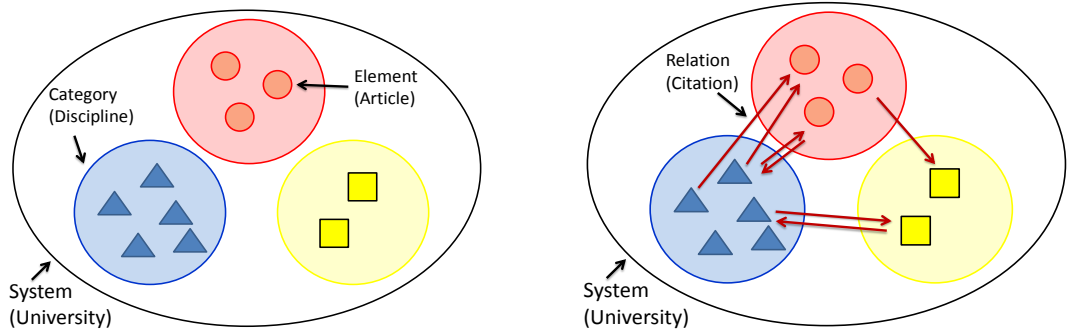

**Figure 1. Illustration of definitions of *diversity* (left) and *coherence* (right).** In parenthesis, an example of the concept: the disciplinary diversity of a university by assigning articles to disciplines, and the disciplinary coherence by means of cross-disciplinary citations. Large circles represent categories. Small figures (triangles, squares and small circles) represent elements.

The novelty and key contribution in Stirling's heuristic for diversity (1998, 2007) is the introduction of a distance metrics $d_{ij}$ between categories. The idea, as illustrated in Figure 2, is that diversity of a system increases not only with more categories (higher *variety*), or with a more balanced distribution (higher *balance*), but also if the elements are allocated to more different categories (higher *disparity*). All other things being equal, there is more diversity in a project including cell biology and sociology than in one including cell biology and biochemistry. While measuring the proportion $p_i$ of elements in a category is straight-forward, providing an estimate of cognitive distance $d_{ij}$ is more challenging. For this purpose, the metrics behind the global maps of science developed in the 2000s have been very useful (Boyack et al., 2005; Moya-Anegón et al., 2007; Klavans and Boyack, 2009; Rafols et al., 2010).

Coherence, on the other hand, aims to capture the extent to which the various parts in the system are directly connected via some relation. For example, the disciplinary coherence of a university (system) can be proxied by the citations (relations) from articles in one WoS category to references in another WoS categories (categories) (Rafols, Leydesdorff et al., 2012). Or it may be explored using network properties at the element level, such as network density or intensity (Rafols and Meyer, 2010).

Further research is needed to establish how and whether coherence can be measured. In this chapter, I tentatively propose that coherence can be thought as having the attributes of density (analogue of variety), intensity (analogue of balance) and disparity, as shown in Figure 3. For this purpose, let me define *M* as the number of existing relations in the systems (out of *N(N-1)* relations possible with *N* categories), intensity of a relation $i_{ij}$ as the scalar representing the relative strength of a relation between categories i and j. Now we can define:

- *density*: number of relations between categories
- *intensity*: overall intensity of the relations in the system.
- *disparity:* degree to which the categories of the relations are different.

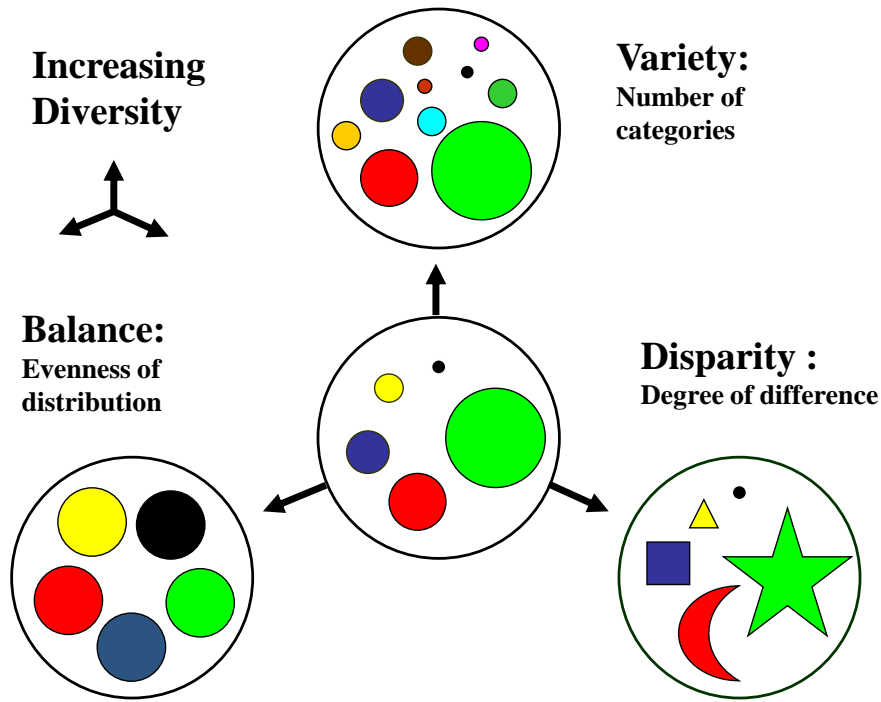

**Figure 2. Schematic representation of the attributes of diversity, based on Stirling (1998, p. 41).** Each full circle represents a *system* under study. The coloured figures inside the circle are the *categories* into which the *elements* are apportioned. Different shapes indicate more difference between categories. Source: Rafols and Meyer (2010).

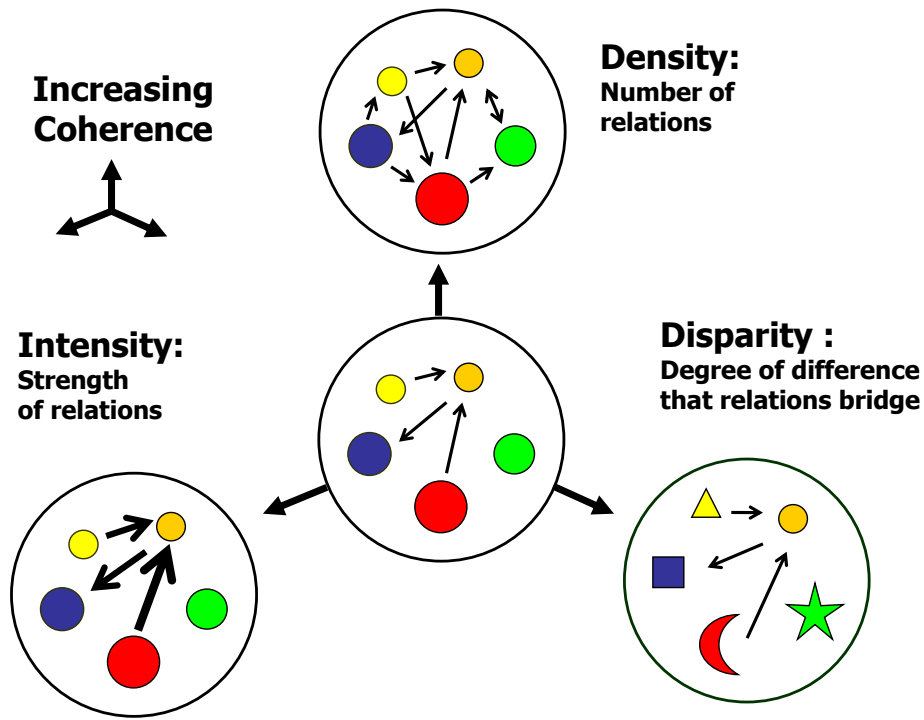

**Figure 3. Schematic representation of the attributes of coherence.** Each circle represents the *system* under study. The coloured figures inside the circle are the *categories* into which the *elements* are apportioned. The lines represent the relations between categories. Thicker lines indicate higher intensity in relations. Different shapes indicate more difference between categories.

Since both diversity and coherence have various aspects, one can generate different, equally legitimate measures of each depending on how these aspects are weighted, as illustrated in Table 1. Stirling (2007) proposed a generalised formulation for diversity which can be turned into specific measures of diversity such variety or the Simpson diversity, by assigning values to the parameters α and β. Ricotta and Szeidl (2006) achieved the same result with a slightly different mathematical formulation (possibly more rigorous but also more cumbersome). In this chapter, I tentatively introduce the same type of generalisation for coherence.

From these considerations, it follows that none of the measures in Table 1 should be taken then as a 'definitive' and 'objective' manner of capturing diversity and coherence. Instead, all measures of diversity and coherence are subjective in the sense that they are derived from judgements about: (i) the choice of categories, (ii) the assignation of elements to categories, (iii) what constitute an adequate metrics of intensity $i_{ij}$, (iv) of a cognitive distance $d_{ij}$ and, finally (v) a judgment regarding what are the useful or meaningful values of α and β for a specific purpose of the study. For example, assuming a distance $0 < d_{ij} < 1$, the analyst would use small values of β to emphasize the importance of distance in the problem under study (this is relevant in issues such as climate change where understandings from social natural sciences need to be integrated). Small values of α, on the contrary, highlight the importance of contributions by tiny proportions. Another possibility is to use various measures of diversity, each of them highlighting one single aspect, as proposed by Yegros-Yegros et al. (2013) (see also in Rafols, Leydesdorff et al. (2012) and Chavarro et al. (2014)). Visualisation in overlay science maps is a way of providing a description of diversity and coherence without need to collapse the data into a single figure (Rafols et al., 2010).

**Table 1. Selected measures of diversity and coherence.** The two comprehensive measures which have been used and tested in the literature are highlighted.

| Notation: | |
|---|---|
| Proportion of elements in category *i*: | $p_i$ |
| Intensity of relations between categories *i* and *j*: | $i_{ij}$ |
| Distance between categories *i* and *j*: | $d_{ij}$ |
| **Diversity Indices:** | |
| Generalised Stirling diversity | $\sum_{i,j(i \neq j)} (p_i p_j)^\alpha d_{ij}^\beta$ |
| Variety (α=0, β=0) | N |
| Simpson diversity (α=1, β=0) | $\sum_{i,j(i \neq j)} p_i p_j = 1 - \sum_i p_i^2$ |
| Rao-Stirling diversity (α=1, β=1) | $\sum_{i,j(i \neq j)} p_i p_j d_{ij}$ |
| **Coherence Indices:** | |
| Generalised Coherence | $\sum_{i,j(i \neq j)} i_{ij}^\gamma d_{ij}^\delta$ |
| Density (γ=0, δ=0) | M |
| Intensity (γ=1, δ=0) | $\sum_{i,j(i \neq j)} i_{ij} = 1 - \sum_i i_{ii}$ |
| Coherence (γ=1, δ=1) | $\sum_{i,j(i \neq j)} i_{ij} d_{ij}$ |

For the sake of parsimony, in practice most applications have used the simplest formulations with α=1 and β=1. This leads to the Rao-Stirling variant of diversity, $\sum_{i,j(i\neq j)} p_i p_j d_{ij}$. This measure had been first proposed by Rao (1982) and has become known in population ecology as quadratic entropy (Ricotta and Szeidl, 2006). It can be interpreted as a distance weighted Simpson diversity (also known as Herfindahl-Hirschman index in economics-related disciplines). Rao-Stirling can be interpreted as the average cognitive distance between elements, as seen from the categorisation, since it weights the cognitive distance $d_{ij}$ over the distribution of elements across categories $p_i$. Similarly, if the intensity of relations is defined as the distribution of relation (i.e. if $i_{ij}=p_{ij}$), the simplest form of coherence, for γ=1 and δ=1, $\sum_{i,j(i\neq j)} p_{ij} d_{ij}$, can be interpreted as the average distance over the distribution of relations $p_{ij}$, rather than the distribution of elements $p_i$.[4]

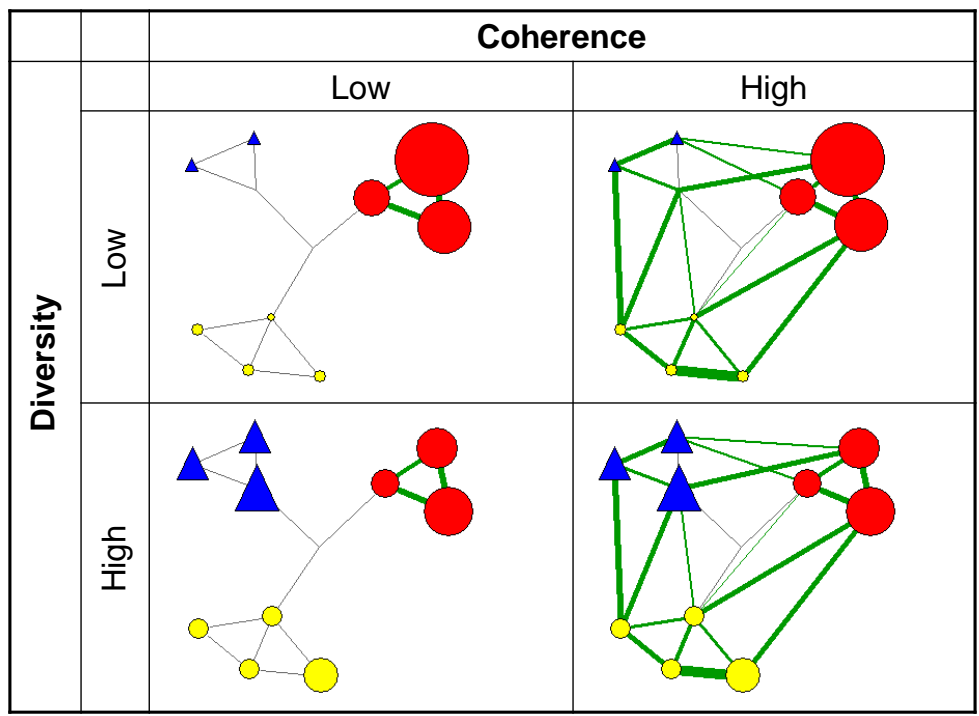

**Figure 4. Conceptualisation of knowledge integration as increase in cognitive diversity and coherence.** Each node in the networks represents a cognitive category. Light grey lines show strong similarity between categories. Same shapes illustrate clusters of similar categories. The size of nodes portrays the proportion of elements in a given category. Dark (or green) lines represent *relations* between categories. Knowledge integration is achieved when an organisation becomes more diverse and establishes more relations between disparate categories. Source: Rafols, Leydesdorff et al. (2012).

The analytical framework proposed understands knowledge integration as an increase in diversity, an increase in coherence, or both. This would means moving from top to bottom, from left to right, or in diagonal from top-left to bottom-right in Figure 4. Similarly, a diffusion process would be seen as an increase in diversity. Higher coherence in diffusion means that as a research topic reaches new areas, it brings them together, whereas lower coherence means that the "topic" is used instrumentally without necessarily linking the new areas.

---

[4] To my knowledge, coherence had only been introduced in this single form, with intensity defined as the proportion of citations between WoS categories $i_{ij}=p_{ij}$. The form of coherence I adopt in this chapter follows from Soós and Kampis (2012) rather than Rafols, Leydesdorff et al. (2012). In the later, coherence, i.e. $\sum_{i,j(i\neq j)} p_{ij} d_{ij}$ was normalised (divided) by Rao-Stirling diversity, i.e. $\sum_{i,j(i\neq j)} p_i p_j d_{ij}$. Such normalisation was useful to remove the correlation between the two variables, but this now seems to me unnecessarily complicated for a general framework.

**3. Choices on data and methods for operationalisation**

The framework presented so far is very general and does not presuppose a commitment to specific data or methods. Now I will operationalise the approach as it was originally developed in order to capture knowledge integration in scientific processes using bibliometric data. Let us first discuss the variety of possible choices in scientometrics regarding the system (unit of analysis), elements, categories and relations to investigate (see Liu et al. (2012) for a previous discussion on these choices).

*3.1. Unit of analysis*

The unit of analysis for measuring diversity can be an article, a researcher, a department or institute, a university or a research topic such as an emergent technology. One thing to notice is that for small units such as articles, diversity can sometimes be interpreted as knowledge integration, without need of further investigating coherence. For example, Alan Porter's work calls *Integration Score* the specific measure of diversity of WoS Categories in the references of an article (Porter et al., 2007, 2008).

One needs to be cautious with choices of units of analysis that involve small numbers, such as article and researcher, because they may not have many elements for the statistical analysis and the resulting measures could be very noisy, particularly when the low numbers are compounded by inaccurate assignation of elements to categories (as it happens when references are assigned to WoS categories). Thus, article or researcher level measures should be treated with caution -- most of the time they will not be reliable individual descriptions, but they can be used averaged over classes --e.g. comparing interdisciplinarity between disciplines using the average disciplinary diversity of references in articles, using samples of some hundred articles (Porter and Rafols, 2009), or to carry out econometric regression models using thousands of articles to investigate the influence of diversity of references on some variables such as number of citations (Yegros-Yegros et al., 2013) or local orientation of the research (Chavarro et al., 2014).

An important consideration in choosing unit of analysis is the recent finding by Cassi et al. (in press) that the Rao-Stirling diversity can be added over scales (under some plausible assumptions, in particular the use of cosine similarity). This means, that the diversity of a research institute is the sum of the diversities *within* each article it published, plus the diversity *between* the articles. This property opens up the possibility of measuring the diversity of large organisations in a modular manner.

*3.2. Classifying elements into categories*

Next, we need to choose the elements 'contained' within the unit of analysis and the categories into which they will be classified. The choice of elements is straightforward. They will typically be articles, references, authors, organisations (as shown in the address or affiliation), or keywords that are listed in the bibliographic record. The challenge is how to classify the elements into categories. Table 2 provides a partial review of different choices of unit of analysis, elements and categories. Since cognitive distance is a key component in the measures of diversity and coherence, the availability of a cognitive metrics among the categories of the classification used is a relevant factor to take into account. For the choice of metrics, see reviews (Börner et al., 2003; Boyack et al., 2011).

In science, disciplines are the most conventional cognitive categories. Most database providers assign articles (usually via journals) to some type of disciplinary categories. Therefore, the most straightforward way of assigning bibliographic elements such as articles or references to categories is to rely on categories provided by databases. The most widely used classification is Thomson-Reuters' Web of Science categories, which is journal-based and very problematic (Rafols and Leydesdorff, 2009), given that articles within a journal do not necessarily share a similar topic or disciplinary perspective.

**Table 2. Examples of different choices of systems, elements, categories and metrics used in measures of diversity.**

| System (unit of analysis) | Elements | Category | Metrics | Examples |
|---|---|---|---|---|
| Article | References in article | WoS Categories | Cosine similarity of WoS Categories | Porter and Rafols, 2009 |
| Article | Citations to article | WoS Categories | Cosine similarity of WoS Categories | Carley and Porter, (2012) |
| Author | Articles | WoS Categories | Cosine similarity of WoS Categories | Porter et al. (2007) |
| University department or Institutes | Articles | WoS Categories | Cosine similarity of WoS Categories | Rafols, Leydesdorff et al. (2012) Soós and Kampis, (2011) |
| Institutes | Articles | 250 Clusters from 300,000 French publications (2007-10) | Cosine similarity of clusters | Jensen and Lutkouskaya (2014) |
| Topic (emergent technology) | Articles | WoS Categories | Cosine similarity of WoS Categories | Leydesdorff and Rafols (2011a) |
| Journals | References in articles of journals | Journals | Cosine similarity of journals | Leydesdorff and Rafols (2011b) |
| Topic (emergent technology) | Articles | Medical Subject Headings (MeSH) | Co-occurrence of MeSH terms in articles | Leydesdorff, Kushnir et al. ( 2012) |
| Topic (emergent technology) | Articles | Medical Subject Headings (MeSH) | Self-Organising Maps based on MeSH, titles, abstracts, references | Skupin et al. (2013) |
| Topic (research) | Patents | Keywords | Self-Organising Maps | Polanco et al. (2001) |
| Open-ended: Topic, Country, Organisation | Patents | International Patent Classification (IPC) Classes | Cosine similarity of IPC classes | Kay et al., (2012); Leydesdorff, Kushnir et al. (2012) |
| Open-ended: Topic, Country, Organisation | Patents | Technological aggregations of IPC classes | Cooccurrence of IPC classes | Schoen et al. (2012) |

Next step is to compute the cognitive distance between categories. As in the case of the classification, the choice of a specific cognitive distance has to be based on judgement. A plausible choice is to take $d_{ij} = (1-s_{ij})$, where $s_{ij}$ is the cosine similarity of the WoS categories. This data is available in excel files at Loet Leydesdorff's website (http://www.leydesdorff.net/overlaytoolkit) from 2007 onwards (Leydesdorff , Carley et al., 2012).

There is, though, the possibility of defining distance in different ways even if you start from the cosine similarity between WoS categories. For example, Soós and Kampis (2012) proposed to define $d_{ij}$ as the sum of the (1-$s_{ij}$) weights of edges in the shortest path from WoS categories *i* to *j*. Jensen and Lutkouskaya (2014) use $d_{ij}=1/s_{ij}$ instead. In these two choices, more weight is given to the co-occurrence of very disparate categories (where the shortest path is long and $1/s_{ij}$ is high) than in the standard similarity. The downside of these alternatives is that the diversity measure is not any more bounded between 0 and 1.

In order to avoid using the journal classifications from data providers (lacking transparency), another possibility is to use journals as categories *per se* to compute diversity measures (Leydesdorff and Rafols, 2011b). The problem here is that most journals are only similar to a small set of related journals. As a result the cosine distances between most journals are practically zero. Since, in principle, the measure of cognitive distance aims to describe differences between distant areas, this result is not useful to capture cognitive distances across disciplines. To overcome this difficulty, Leydesdorff , Rafols et al., (2013) have recently proposed to use the distance observed in the two dimensional projection of a map of the +10,000-dimensions of the actual distance matrix. This is a very coarse approximation, but has the advantage of distributing quite evenly the distances among journals between zero and a maximum value (which we redefine as one).

Another alternative to the inaccuracies of WoS or Scopus categories to is to carry out clustering using bibliometric data (Rafols and Leydesdorff, 2009). These bottom-up categories may be more consistent with research practices, at least as seen from citation patterns. They can be based on journal clustering (e.g. more than 500 categories in the UCSD by Börner et al. (2012)[5], see also Rosvall and Bergstrom, (2008)) or on paper-level clustering (e.g. about 700 categories by Waltman and van Eck, 2012). [6]

In all this analysis so far, we have relied on static classifications with stable categories of all science.[7] In the case of knowledge integration, this is useful to characterise the knowledge background from a traditional perspective of science (e.g. in terms of subdisciplines). In the case of emergent technologies, though, new research topics do not conform to these traditional categories and it may be illuminating to complement the traditional view with a more fine-grained, local, bottom-up and dynamic classification. The difficulty of this approach is that constructing very fine-grained and/or dynamic clusters that are meaningful is very demanding (Havemann et al., 2012). Since noise increases as the sample becomes smaller, many clusters become unstable (are born, die, divide, etcetera) below a threshold around 100-1,000 papers (Boyack et al., 2013), and their local structure may differ from the one obtained with a global map (Klavans and Boyack, 2011) . This clustering has been the approach of Kajikawa and colleagues, using direct-citation-link clustering, for example in studies on energy technologies (Kajikawa et al., 2008) or bionanotechnology (Takeda et al., 2009). Boyack et al. (2013) are also following this approach with very small clusters. In principle, the framework proposed here might also work with small and dynamic categories --in practice, the challenge is constructing these categories.

Rather than relying on aggregate categories, one may try to use directly the elements as categories calculating their cognitive distance without further categorisation, as a contrast to the coarse-grained, static classification (Rafols and Meyer, 2010; Soós and Kampis, 2011). Jensen and Lutkouskaya (2014) use various measures of diversity with different categorisations in order to have a more plural view of the degree of interdisciplinarity of French national laboratories. These efforts align with the conceptualisation of scientometric advise as helping the opening-up of perspectives in science policy debate, rather than narrowing the scope of decisions (i.e. closing-down) (Barré, 2010; Rafols, Ciarli et al., 2012).

---

[5] This classification and underlying map can be downloaded and publicly used. It is available at http://sci.cns.iu.edu/ucsdmap/.
[6] This classification is available at http://www.ludowaltman.nl/classification_system/
[7] According to Boyack et al., (2013), more than 99% of clusters are stable at a level of aggregation of about 500 clusters for all science.

Finally, instead of using classifications that relate bibliometric elements with a cognitive category based on scientific point of view such as a subdiscipline, an emergent fields or a research topic, as discussed above, one may instead relate the elements with categories from outside science such as diseases or technologies. The Medical Subject Headings (MeSH) of PubMed offer a way of making the linkages between elements of a publication and the specific practitioner-oriented perspectives of its hierarchical classification, such as descriptors for disease, techniques and equipment, chemicals/drugs, and healthcare. Using one or more of these practitioner-oriented categories might be particularly helpful when analysing the social impact of research. Leydesdorff, Rotolo et al. (2012) and Skupin et al., (2013) have recently created global MeSH maps. However, unlike the globals maps of science, which show consensus (Klavans and Boyack, 2009), these maps could not be matched. Hence, I would suggest that the underlying cognitive structure and metrics of MeSH deserve further investigation.

*3.3. Capturing relations*

In order to measure coherence one needs to associate relations observed in the system with links among categories. Since these relations are derived from information within or between elements, the discussion in the previous subsection on the assignation elements into categories is directly applicable to relations as well. For example, a citation allows us to relate the category of an article to the category of one its references. The challenge, as discussed, is how the article and the reference are classified into WoS categories, journals, bottom-up clusters, or MeSH terms, etc. Another straightforward way to create relations is from co-occurrences of some article attributes. For example, if MeSH terms are taken as categories, the strength of the relation between two MeSH can be estimated as their normalised number of shared publications.

An interesting point to notice regarding relations is that they do not need to be symmetrical, i.e. $i_{ij} \neq i_{ji}$. This is obvious for directed flows: it is well known, for example, that an applied research field like oncology cites cell biology proportionally more than the reverse (4.5% vs. 7.5% citations in 2009). In the case in which relations are non-directed (i.e. edges), such as co-occurrences, it is also possible to do an asymmetrical normalisation, i.e. to normalise $i_{ij}$ according to counts in $i$ category only. This raises the interesting question of whether cognitive distances, which in most studies are symmetrical ($d_{ij}=d_{ji}$), should also be taken as asymmetrical -- an issue which deserves a full separate discussion.

*3.4. Visualisation*

Given that diversity and coherence are multidimensional concepts, visualisation can be helpful to intuitively present the various aspects without collapsing all the information into a single value. The method proposed here relies on the ideas of *overlaying* (projecting) the elements of the unit of analysis over the cognitive space selected, an idea that I borrowed from Kevin Boyack and colleagues (Boyack et al., 2009). The visualisation has three steps. First, one builds a 'basemap' representing the cognitive space selected. A widely used 'basemap' is the global map of science, representing the disciplinary structure of science (freely available at Rafols et al., 2010). The map intuitively portrays the cognitive distance between its nodes, the WoS categories (or others).

Second, one projects the distribution of elements into categories over the basemap by making the size of each category (node) proportional to the frequency of elements in that specific category. This means for example that the size of a node in the global map of science is made proportional to the number of articles published in that WoS category in the sample studied. This projection or overlay allows the viewer to capture intuitively the three attributes of diversity: First, the map captures *variety* by portraying the number of categories in which a unit of analysis (e.g. university) is engaged. Second, it captures *balance* by presenting the nodes with different sizes. Third, unlike bar charts, the map conveys *disparity* among categories by illustrating the cognitive distance by means of the physical distance in the map (Rafols et al., 2010, p. 1883).

The third step is to project the relations over the map as illustrated in Figure 5 (Rafols, Leydesdorff et al., 2012). This projection is perhaps the most unconventional step, since it consists of overlaying the links in the unit of analysis, over the structure of the global map, without re-positioning the nodes. The intensity of the relations is shown by the thickness of the links. It is precisely the contrast between the local relations (in thick darker lines) in comparison to the global relations (in finer lighter lines) what allows us to understand the nature and extent of knowledge integration that is being carried out. The visualisation of relations between hitherto unrelated bodies of knowledge conveys intuitively the concept of coherence.

The maps shows intuitively the three aspects of coherence: whether coherence is achieved across many categories (density), the thickness of links (intensity) and whether they are linked across distant categories (disparity). Since the probability of links does not only depend on cognitive proximity, it is useful to make an overlay of the expected relations (in the case of citation, this depends both on citation sources and probability flows) and one overlay of the observed relations, as shown in Figure 5.

**4. How-to compute and visualise knowledge integration**

This section describes the protocol of the method to compute and visualise diversity and coherence. To do so, I will follow the most well-established application of this framework, based on the so-called global maps of science based on WoS categories. Since these categories are not very accurate, it is best to think this analysis as merely exploratory or illustrative. Detailed information on this method is presented in the Annex of Rafols et al. (2010). The data and basemaps used here are publicly available at Loet Leydesdorff's website http://www.leydesdorff.net/overlaytoolkit.

*4.1 Illustrative introduction to measures of diversity*

This protocol illustrates how to compute diversity and coherence using excel files and Pajek maps.. Supplementary files are available here:   http://www.sussex.ac.uk/Users/ir28/book/excelmaps

Data collection
1. Delineate and download the data set from the Web of Science.

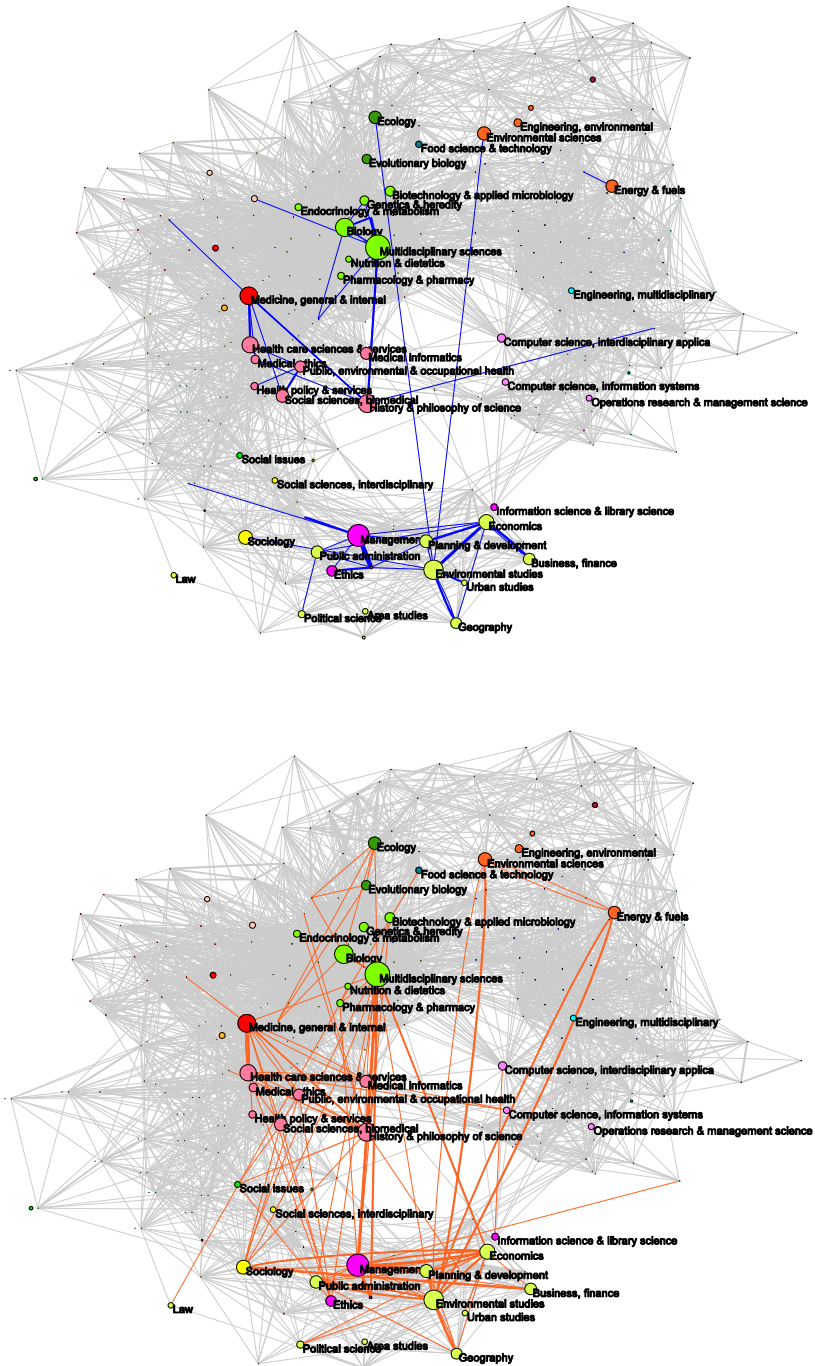

**Figure 5. Expected (top) and observed (bottom) citations of the research centre ISSTI (University of Edinburgh) across different Web of Science categories.** The grey lines in the background show the global map of science (Rafols et al., 2010). The size of the nodes reflects the aggregate number of citations given to a field from all ISSTI's publications. Blue lines show the expected citations between fields, given where ISSTI is publishing. The computation of expected citations is based on the number of publications in a field, and the average proportion of citations to other fields in all the WoS. It can be observed that the expected citations tend to be within disciplines: within biological sciences, within health services, and within social sciences. Orange lines show the citations between fields observed in ISSTI's publications. The citations between fields criss-cross the map of science both within disciplines and across disciplines. (Only citations larger than 0.2% of ISSTI's total are shown). Source: Rafols, Leydesdorff et al. (2012).

Measure of diversity

For the sake of helping non-expert readers, the measures are presented in the spread-sheet calculations.

2. **Create a list** with the distribution of WoS categories. These are listed in the field "WC" in the file downloaded.
3. **Open** the spread sheet file "DiversityComputation2009.xlsx".
4. **Paste** the list in the tab "INPUT". Notice that only the WoS Categories in the Journal Citation Index in 2009 are present. Other categories will not be counted.
5. **Go to tab** "OUTPUT". **Select a threshold** for the minimum proportion to be taken into account in counting variety and disparity (default = 0.01, i.e. 1%)
6. The file provides values for Rao-Stirling diversity and other measures of diversity as described in Table 1.

Measure of coherence

7. **Create a matrix** with the ordered distribution of citations from the WoS categories to WoS categories in the data set. (Unfortunately, to my knowledge this cannot be done with publicly available software. VantagePoint[8] provides an easy template to create it)
8. **Open** the spread sheet file "CoherenceComputation2009.xlsx".
9. **Paste** the matrix in the tab "INPUT matrix".
10. **Go to tab "OUTPUT"** to retrieve the data on coherence.

Visualisation of diversity with Pajek
11. **Open** the Pajek file "ScienceMap2009.paj" (press F1)
12. **Upload** the vector file (.vec) with the distribution list of WoS categories "ListWoSCats.vec"
**Press Ctrl-Q** to visualise the overlay map (details provided in the appendix of Rafols et al. (2010)).

*4.2 R script for computing diversity of a set of articles*

This protocol provides a script for computing diversity over large data sets. Supplementary files are available here:  http://www.sussex.ac.uk/Users/ir28/book/diversity.zip

The file "diversity_measures_1.R" contains the script with the programming language R to compute the Rao-Stirling diversity for each individual article of a list of articles, based on the assignation of references to WoS categories. It requires the file with the proximity matrix ("cosine_similarity_matrix_sc.csv") and an input file with the list distribution of WoS in the reference list, as shown in "articles_sample.csv").  The directory with the file needs to be written up into the script before running it.

**5. Conclusions**

In this chapter I have presented a framework for the analysis of knowledge integration and diffusion based on the concepts of cognitive diversity and coherence. Knowledge diffusion is seen as an increase in the cognitive diversity of the areas to which a given discovery or technology has spread. Knowledge integration, is seen as an increase in cognitive diversity and/or coherence. The chapter introduced the general mathematical formulation of these concepts. It has proposed that diversity has the attributes of variety, balance and disparity, whereas and coherence has the attributes of density, intensity and disparity. Diversity and coherence can be formulated in various manners depending on the relative weight of the attributes –hence their values will depend on the choice of weight given to them.

---
[8] http://www.thevantagepoint.com/

Given the importance of the choices of elements, relations and classifications to characterise diversity, I have discussed the different approaches to classify science into categories, from the top-down and coarse classifications such as the WoS' to more fine-grained categories. I have briefly mentioned the possibilities of characterising science with more practitioner-oriented perspectives such as those provided by MeSH terms. I have illustrated with a spread sheet how to compute diversity and coherence using WoS categories. Since WoS categories are very inaccurate, this method should be interpreted as exploratory.

The fact that diversity and coherence can be measured using various mathematical formulations and that, for each of them, various operationalisations are possible in terms of the elements and categories chosen, should send a strong message of caution: knowledge integration and diffusion are strongly dependent on the perspective taken. It could be that with a disciplinary perspective, a research topic has become stagnant (staying within the same discipline), but with a medical perspective, the topic is diffusing to new areas such as new diseases. Hence, the measures and maps should be read as inevitably partial perspectives -- not covering but a few of the possibilities for capturing knowledge dynamics. Other dynamics of knowledge integration, not covered by diversity and coherence are also possible. For example, 'intermediation' would be another way to capture knowledge integration focussing in the bridging processes (Chen et al., 2009; Rafols, Leydesdorff et al., 2012).

The framework proposed has been developed for mapping in the conventional cognitive dimension of science (disciplines and topics), but it can easily be extended to other cognitive perspectives such as those arising from medicine (via MeSH). Similarly, the approach can be easily extended to patents, using global maps of technology (Kay et al., 2012; Leydesdorff, Kushnir et al., 2012; Schoen et al., 2012), and closely related measures of diversity (Nesta and Saviotti, 2005, 2006).

Finally, I would like to highlight that while the framework has been applied to cognitive distance, it can in principle be applied as well to other analytical dimensions. For example, one might look at the geographical diversity of a collaborative project not counting the number of countries, but investigating collaborations or citations in terms of geographical distance (Ahlgren et al., 2013). Or investigate the diversity in organisations in a new topic not just counting organisations, but taking account the cognitive proximity of the organisations. As proposed by Frenken (2010), by extending this framework to other analytical dimensions, it would be possible to investigate how knowledge integration is mediated by geographical, organisational, institutional and social networks.

**Acknowledgements**

This chapter summarises work carried out with many collaborators, in particular with L. Leydesdorff, A.L. Porter and A. Stirling. I am grateful to D. Chavarro for writing the code in R language to compute diversity. I thank Y.X. Liu, R. Rousseau and A. Stirling for fruitful comments. I acknowledge support from the UK ESRC grant RES-360-25-0076 ("Mapping the dynamics of emergent technologies") and the US National Science Foundation (Award #1064146 - "Revealing Innovation Pathways: Hybrid Science Maps for